\documentclass[11pt,twoside]{article}

\usepackage{asp2004}
\usepackage{epsf}
\usepackage{psfig}
\usepackage{lscape}
\usepackage{graphicx}

\newcommand{\Mstar}{\ensuremath{M_{\star}}}

\begin{document}
\title{Clusters of Galaxies at $1 < z< 2$ : The Spitzer Adaptation of the Red-Sequence Cluster Survey}   
\author{Gillian Wilson\altaffilmark{1}, Adam Muzzin\altaffilmark{2}, Mark Lacy\altaffilmark{1}, Howard Yee\altaffilmark{2}, Jason Surace\altaffilmark{1}, Carol Lonsdale\altaffilmark{3}, Henk Hoekstra\altaffilmark{4}, Subhabrata Majumdar\altaffilmark{2}, David Gilbank\altaffilmark{2} \& Mike Gladders\altaffilmark{5}  (``The SpARCS Collaboration'')}   

\altaffiltext{1}{Spitzer Science Center, California Institute of Technology, 220-6, Pasadena, CA 91125; gillian@ipac.caltech.edu}

\altaffiltext{2}{University of Toronto, Department of Astronomy \& Astrophysics, 60 St. George St, Toronto, ON, M5S 3H8, Canada}

\altaffiltext{2}{Infrared Processing and Analysis Center (IPAC), California Institute of Technology
MS 100-22, 770 South Wilson Avenue, Pasadena, CA 91125}

\altaffiltext{4}{University of Victoria, Elliott Building, 3800 Finnerty Rd,  Victoria, BC, V8P 5C2, Canada}
\altaffiltext{5}{Carnegie Observatories, 813 Santa Barbara Street, Pasadena, California 91101}

\begin{abstract} 

As the densest galaxy environments in the universe, clusters are vital to
our understanding of the role that environment plays in galaxy
formation and evolution.  Unfortunately, the evolution of
high-redshift cluster galaxies is poorly understood because of the
``cluster desert'' that exists at $1 < z <2$. 
The SpARCS collaboration is currently 
carrying out a 1-passband ($z^\prime$)  imaging survey which,
when combined with the pre-existing  $\sim 50$ deg$^2$ $3.6 \micron$ Spitzer SWIRE Legacy
Survey data, will efficiently detect hundreds of clusters in the cluster
desert using an infrared application of the well-proven cluster
red-sequence technique.  We have already tested this 1-color 
(z$^\prime - [3.6]$) approach using a 6 deg$^2$
``pilot patch'' and shown it to be extremely
successful at detecting clusters at $1 < z < 2$.  The clusters
discovered in this project will be the first large sample of
``nascent'' galaxy clusters which connect the star-forming proto-cluster
regions at $z >$ 2 to the quiescent population at $z < 1$.
The existing seven-passband Spitzer data (3.6, 4.5, 5.8,
8.0, 24, 70, 160 $\micron$) will allow us to make the first
measurements of the evolution of the cluster red-sequence, IR
luminosity function, and the mid-IR dust-obscured star-formation rate 
for $1< z <2$ clusters.
\end{abstract}

\section{Introduction}   
Galaxy clusters are unique, high-density regions
in the universe and are therefore important testbeds for theories of
galaxy formation and evolution.  Within the last decade, significant
amounts of observational resources have been invested in finding
clusters to $z \sim$ 1 using X-ray \citep{ros98, p04} and optical \citep{gy00, gil04}
techniques.  Simultaneously, surveys which employ
the Lyman-break technique have begun to uncover a population of
vigorously star-forming proto-cluster regions at  $2< z < 5$ (e.g., \citealt{kurk04, ouchi05, steidel05}).  Tantalizingly, the crucial stages in cluster galaxy
evolution, where star-forming Lyman-$\alpha$ galaxies transform into
the quiescent early-type population seen at $z \sim$ 1 \citep{blake03, hold04}
appear to occur in the hard to access redshift range $1 < z <2$,
the ``cluster desert''.  Currently, there are fewer than 15
spectroscopically-confirmed clusters  at $1 < z < 2$\footnote{NASA Extragalactic Database (NED)} (with the highest-redshift being
$z = 1.41$; \citealt{stan05})

\section{The (Infrared) Cluster Red-Sequence Method}	

The Cluster Red-Sequence (CRS) technique  requires imaging in only two filters which span the 
$4000 \AA$ break feature in early types, and is a well-tested and observationally efficient 
method for detecting clusters \citep{gy05, w05}. 
The CRS algorithm selects clusters by using the fact that 
the cluster galaxy population is dominated by ellipticals and that they
lie along a linear relation in the color-magnitude plane.  If two filters
that span the rest-frame $4000 \AA$ break are used to construct the color-magnitude
diagram, cluster ellipticals are always the brightest, reddest galaxies at
a given redshift, and therefore provide significant contrast with the 
field.  The original technique was first developed by Gladders \& Yee and
is being used very successfully to target clusters in the $0.2 < z <  1.1$
range where the blue-red $R - z^{\prime}$ filter combination spans the $4000 \AA$ break.

Fig.~1 shows the observed spectral energy distribution (SED) of an early type at $z=0.0, 1.0, 1.5$ and 
$2.0$.
Also shown are the portions of the SED sampled by the
$r^{\prime}$, $i^{\prime}$, $z^{\prime}$, $K$ and [3.6] passbands.
At $z \sim 1.1$, the 4000\AA-break is shifted into the infrared (IR).
To carry out a cluster survey at $1 < z < 2$ requires deep IR observations over a large 
area - observations that are very difficult to obtain from the ground.

The only deep, large area (tens of square degrees) IR dataset currently available 
(or indeed available for the foreseeable future)
is the Spitzer SWIRE Legacy Survey\footnote{http://swire.ipac.caltech.edu/swire/swire.html}.

\section{The SpARCS Survey}	

The SWIRE Survey is comprised of six fields, and is well-suited to searching for high 
redshift clusters. 
Four fields (34.2 deg$^2$) are accessible from the North,
and two fields (14.8 deg$^2$) from the South (totalling 49 deg$^2$).
Rich clusters are very rare so it is
important to search over as large an area as possible.
We expect to find only about 30 \emph{rich} clusters (Abell Class 1) in the \emph{entire} 
SWIRE survey.

Our collaboration, SpARCS\footnote{http://spider.ipac.caltech.edu/staff/gillian/SpARCS} (The \emph{Sp}itzer \emph{A}daptation of the \emph{R}ed-Sequence \emph{C}luster \emph{S}urvey) 
has developed  an IR adaptation of the two-filter 
cluster red-sequence technique 
which utilizes IRAC's  $3.6 \micron$ channel as the ``red'' filter
\citep{faz04}. A good choice of
``blue'' filter (better than e.g., $R$ or $i^{\prime}$), is the $z^{\prime}$ filter. This is
because  $z^{\prime}$ samples blueward of rest-frame $4000 \AA$ at $z > 1.1$, and
yet becomes only as blue as rest-frame $U$ at $z=2$.

Fig.~2 shows red-sequence models constructed using the code
of \cite{bc03}, assuming a galaxy undergoing a burst of star-formation at $z_f = 4.0$ and evolving 
passively thereafter. Based on these models, we can detect
\Mstar\ early types to $z\sim 2$.

To date, we have combined 6 deg$^2$ of MegaCam $z^\prime$ and SWIRE [3.6]
data in the XMM-LSS field, and detected about 70 moderately-rich clusters
at $0.05 < z< 1.85$. 
Figs.~3~\&~4 show examples of clusters at $z>1$. We are currently
extending this analysis to the remaining SWIRE fields.


\section*{Science Goals}	

By searching for clusters over 50 deg$^2$,
our aim is to detect a \emph{representative} sample of 
clusters at $1 < z< 2$ . This will allow the study of cluster galaxy
evolution over a range of cluster richnesses and redshifts.

We plan to  measure the IR luminosity function (e.g., \citealt{m05, m06}), the
mid-IR dust-obscured star-formation rate (using the SWIRE $24 \micron$
data), the red-sequence, and the relative color distribution.
In particular, the evolution of the color distribution will
allow us to study the star-formation properties of the cluster galaxies as a
function of mass and redshift and understand how the star-forming
proto-cluster regions at $z > 2$ connect to the quiescent clusters at 
$z < 1$.
Furthermore, by measuring the
evolution of both the IR luminosity function and the slope, scatter
and color of the red-sequence to higher redshift than previous studies, 
we hope to put precise
constraints on the epoch when stellar mass is assembled in cluster galaxies
 \citep{hold04, kod04}.

Our goal is to release our cluster catalogs to the community as soon as
possible, so that this sample can serve as a basis for numerous other
science projects which require follow-up observations (e.g. mass
determination, enrichment history).

\begin{figure}[ht]
  \begin{center}
    \plotone{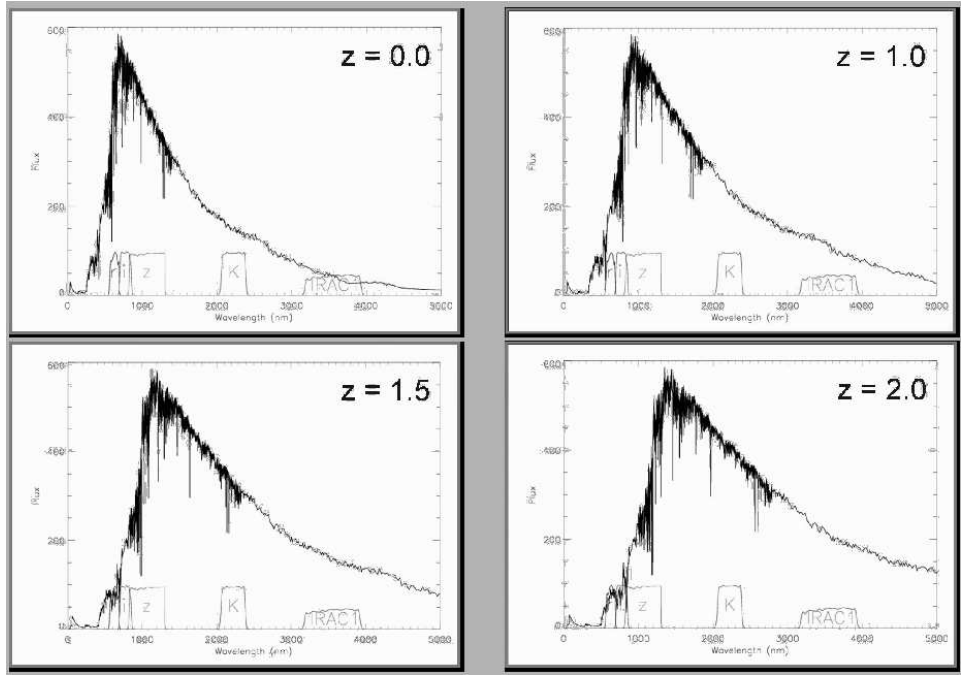}
	\end{center}
\caption{Observed SED of an early type at various redshifts.
At $z~\sim~1.1$, the traditional ``red'' z$^\prime$ filter no longer samples redward of the 
rest-frame $4000 \AA$ break. To detect clusters at higher redshift, an IR ``red'' filter is necessary.
$z^{\prime}$, however, is a good choice of  ``blue'' filter for high redshift cluster surveys.
}
\end{figure}

\begin{figure}[ht]
\begin{center}
    \plotone{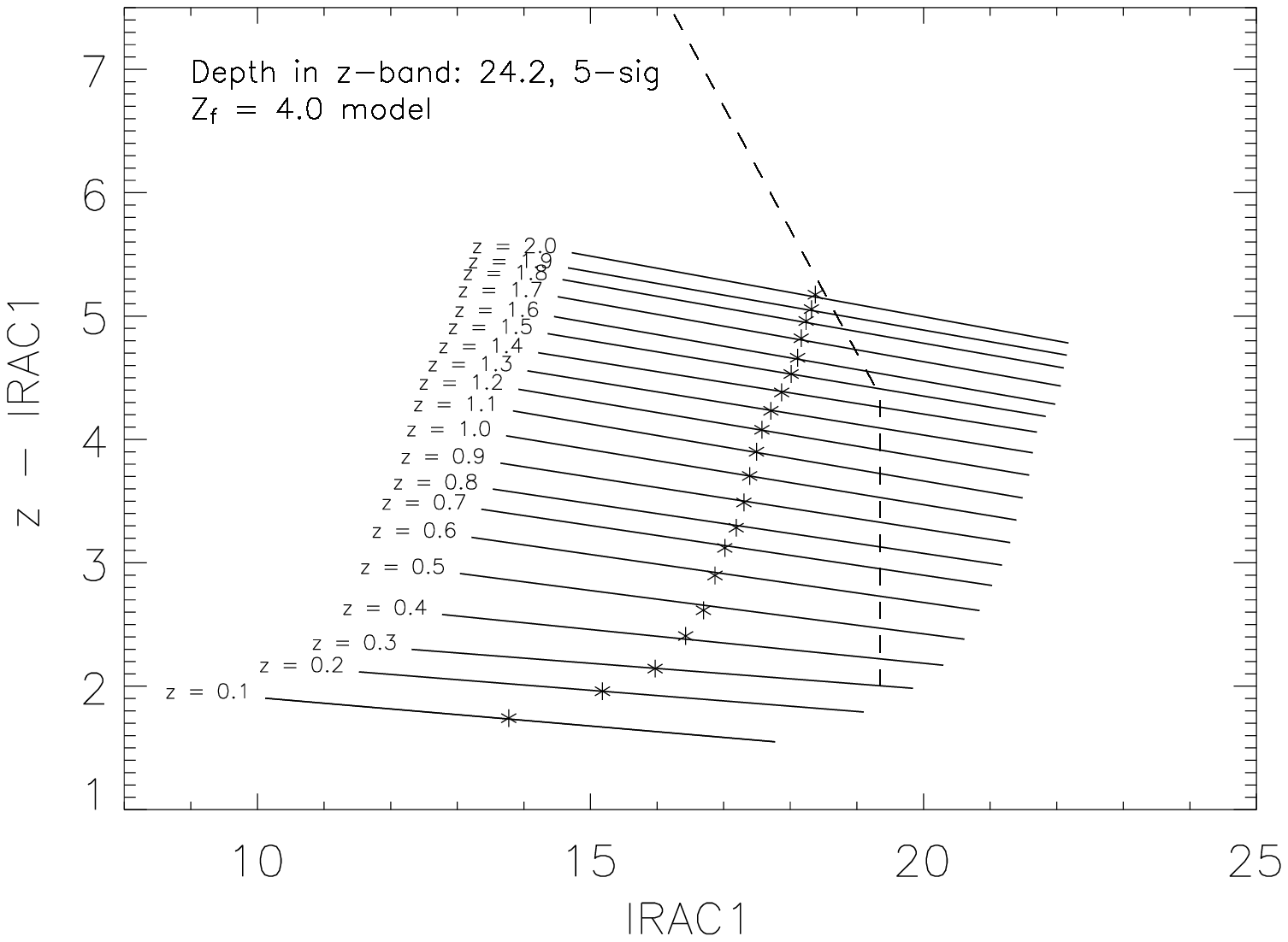}
  \end{center}
\caption{Simulated red-sequences for a z$^\prime$-[3.6] filter combination assuming a \citet{bc03} model
where a galaxy undergoes a burst of star-formation at $z_f = 4.0$ and evolves passively thereafter. The 
dashed line is the depth of our survey. \Mstar\, the characteristic magnitude at each redshift
is denoted by an asterisk. Photometric redshifts can be determined from the red-sequence 
(z$^\prime$ - [3.6] color) to an accuracy of $10 \%$.
}
  \begin{center}
	\plotone{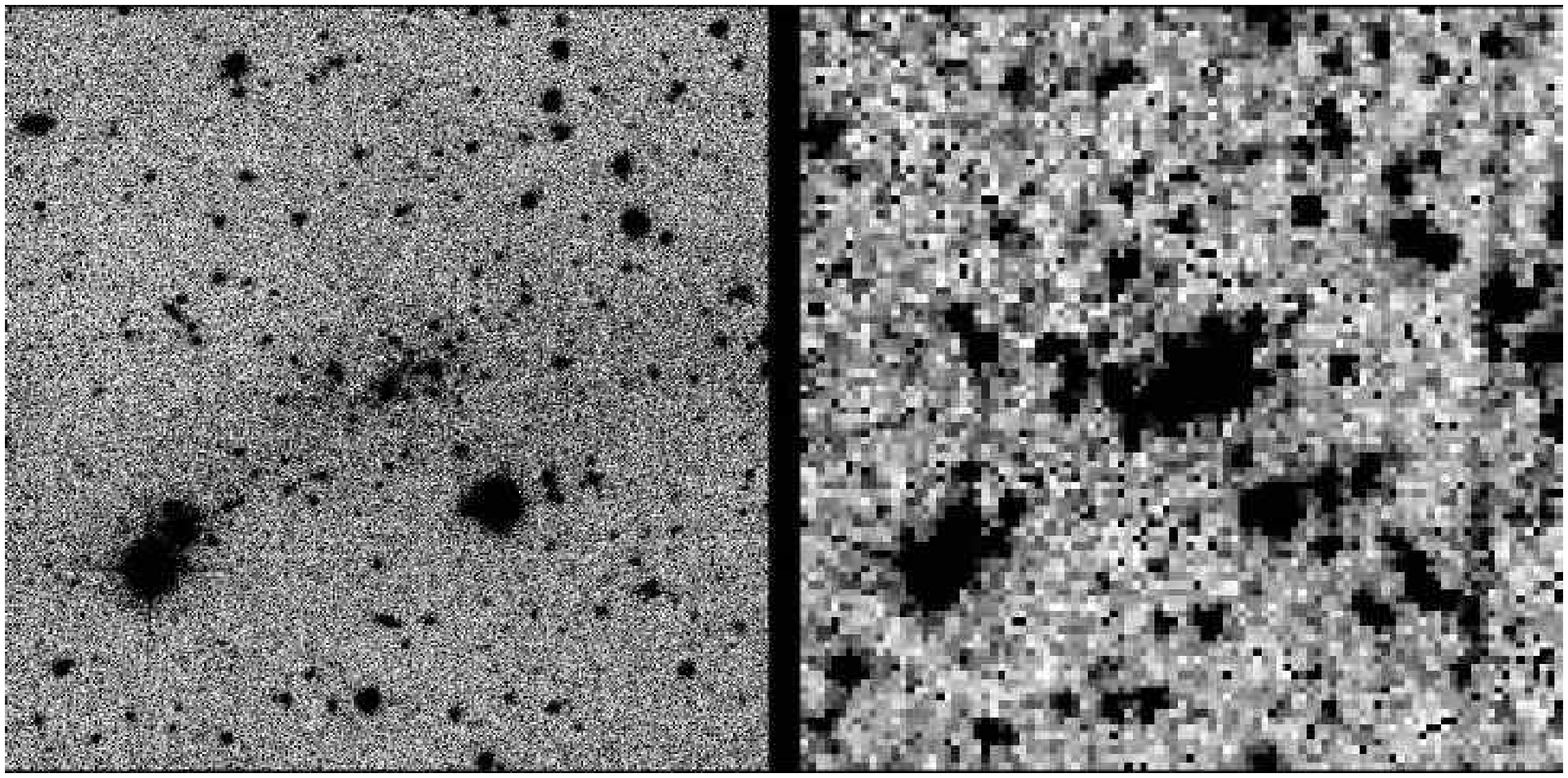}
  \end{center}
\caption{$z^{\prime}$ (left) and [3.6] (right) image of a $z= 1.22$ cluster
in the  6 deg$^2$ ``pilot patch'' (the redshift is estimated
from the color). The FOV is 1 Mpc at the cluster center. For a color image and
more examples see (http://spider.ipac.caltech.edu/staff/gillian/SpARCS).
Note: This cluster was discovered 
independently by the XMM-Newton 
Large Scale Structure Survey \citep{p04}.}
\end{figure}

\begin{figure}[ht]
\begin{center}
    \includegraphics[scale = 1.2]{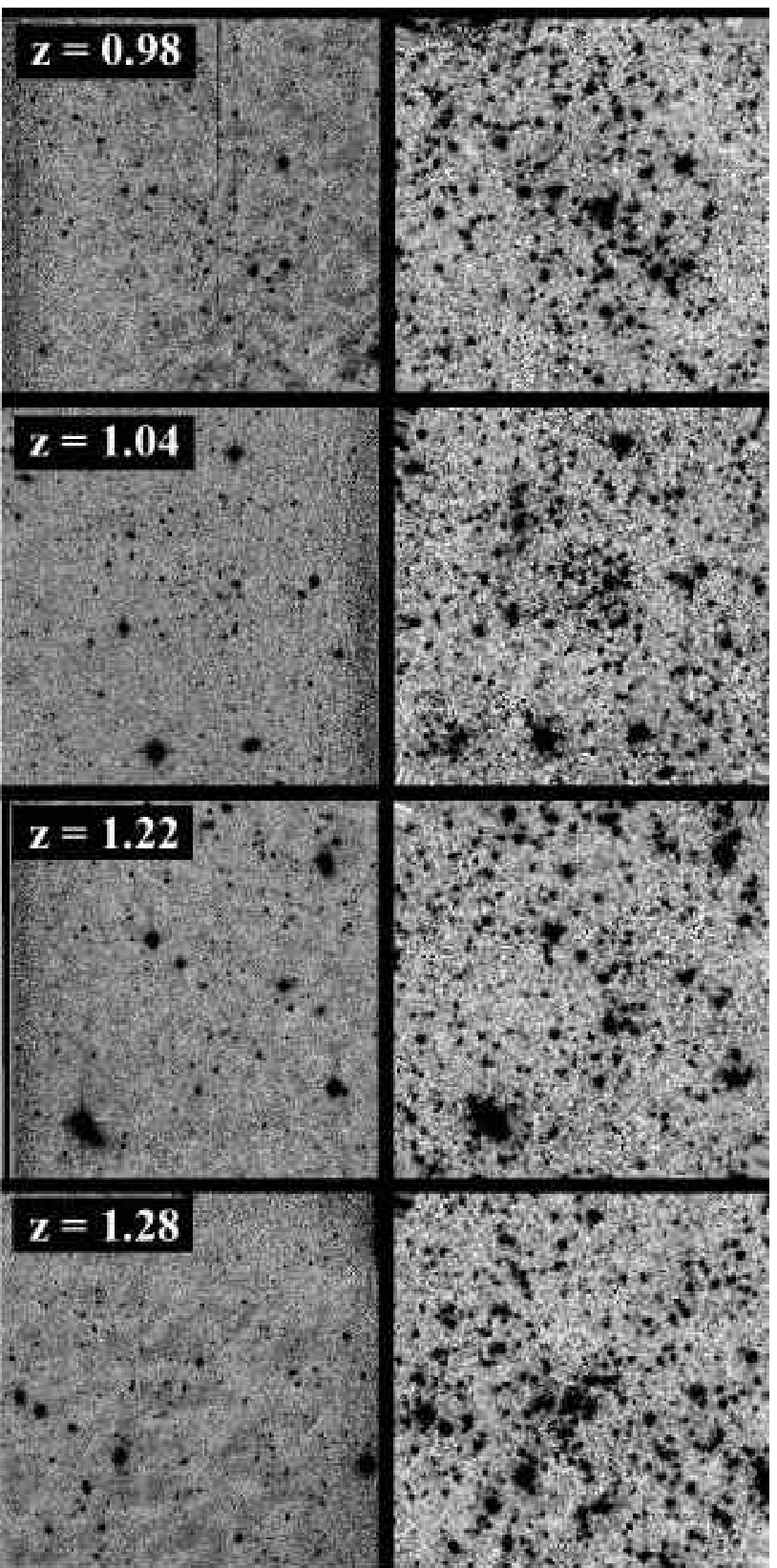}
  \end{center}
\caption{As for Fig.~3 but for a 2 Mpc FOV.
}
\end{figure}

\acknowledgements 

This work is based in part on archival data obtained with the Spitzer Space Telescope, which is operated by the Jet Propulsion Laboratory, California Institute of Technology under a contract with NASA. Support for this work was provided by NASA. This work was also based on observations obtained with MegaPrime/MegaCam, a joint project of CFHT and CEA/DAPNIA, at the Canada-France-Hawaii Telescope (CFHT) which is operated by the National Research Council (NRC) of Canada, the Institut National des Science de l'Univers of the Centre National de la Recherche Scientifique (CNRS) of France, and the University of Hawaii. This work is based in part on data products produced at TERAPIX and the Canadian Astronomy Data Centre as part of the Canada-France-Hawaii Telescope Legacy Survey, a collaborative project of NRC and CNRS.



\begin{thebibliography}{}
\bibitem[Blakeslee et al.\ (2003)]{blake03}
Blakeslee, J. P., et al. 2003, ApJ, 596, 143
\bibitem[Bruzual \& Charlot (2003)]{bc03}
Bruzual, G., \& Charlot, S. 2003, MNRAS, 344, 1000
\bibitem[Fazio et al.\ (2004)]{faz04}
Fazio, G.~G., et al. 2004, ApJS, 154, L10 
\bibitem[Gilbank et al.\ (2004)]{gil04}
Gilbank, D. G., et al. 2004, MNRAS, 348, 551 
\bibitem[Gladders \& Yee (2000)]{gy00}
Gladders, M. D., Yee, H. K. C., 2000, AJ, 120, 2148
\bibitem[Gladders \& Yee (2005)]{gy05}
Gladders, M. D., Yee, H. K. C., 2005, ApJS, 157, 1
\bibitem[Holden et al.\ (2004)]{hold04}
Holden, B. P. et al. 2004, AJ, 127, 2484
\bibitem[Kodama et al.\ (2004)]{kod04}
Kodama, T. et al. 2004, MNRAS, 250, 1005
\bibitem[Kurk et al.\  (2004)]{kurk04}
Kurk, J. D., et al. 2004, A\&A, 428, 817
\bibitem[Muzzin, Wilson \& Lacy (2005)]{m05}
Muzzin, A., Wilson, G., \& Lacy, M. 2005, to appear in ``Spitzer Space Telescope: New Views of the Universe'', ed. Armus L., (astro-ph/0503640)
\bibitem[Muzzin, Wilson \& Lacy (2006)]{m06}
Muzzin, A., Wilson, G., \& Lacy, M. 2006, in prep. 
\bibitem[Ouchi et al.\ (2005)]{ouchi05}
Ouchi, M., et al. 2005, 620, L1
\bibitem[Pierre et al.\ (2004)]{p04}
Pierre, M. et al. 2004, JCAP, 9, 11
\bibitem[Rosati et al.\ (1998)]{ros98}
Rosati, P. et al. 1998, ApJL, 492, L21
\bibitem[Stanford et al.\ (2005)]{stan05}
Stanford, S. A., et al., 2005, ApJ, 634, L129
\bibitem[Steidel et al.\ (2005)]{steidel05}
Steidel, C. C., et al. 2005, ApJ, 626, 44
\bibitem[Wilson, Muzzin \& Lacy (2005)]{w05}
Wilson, G., Muzzin, A., Lacy, M. 2005, to appear in ``Spitzer Space Telescope: New Views of the Universe'', ed. Armus L., (astro-ph/0503638)
\end{thebibliography}
\end{document}